\documentclass[final,5p,times,twocolumn]{elsarticle} 

\usepackage{amssymb}
\usepackage{amsmath}

\usepackage{siunitx}
\usepackage{subcaption}
\usepackage[colorlinks=true, linkcolor=blue, citecolor=blue, urlcolor=blue]{hyperref}
\usepackage[capitalize,nameinlink]{cleveref}
\crefname{figure}{Fig.}{Figs.}

\usepackage{lineno}

\begin{document}

\begin{frontmatter}



\title{Investigating irradiation effects and space charge sign inversion in n-type Low Gain Avalanche Detectors}

\author[label1,label2]{Veronika Kraus \corref{cor1}}
\author[label3]{Margarita Biveinytė}
\author[label4]{Marcos Fernandez Garcia}
\author[label5]{Salvador Hidalgo}
\author[label1]{Michael Moll}
\author[label5]{Jairo Villegas}
\author[label1]{Moritz Wiehe}

\cortext[cor1]{Corresponding author. Email: \href{mailto:veronika.kraus@cern.ch}{veronika.kraus@cern.ch}}

\affiliation[label1]{organization={CERN, Organisation europénne pour la recherche nucléaire},
            addressline={Espl. des Particules 1},
            city={Genève},
            postcode={1217},
            country={Switzerland}}

\affiliation[label2]{organization={TU Wien, Faculty of Physics},
            addressline={Wiedner Hauptstraße 8-10},
            city={Vienna},
            postcode={1040},
            country={Austria}}

\affiliation[label3]{organization={Vilnius University},
            addressline={Universiteto St. 3},
            city={Vilnius},
            postcode={01513},
            country={Lithuania}}

\affiliation[label4]{organization={Instituto de Física de Cantabria, IFCA (CSIC-UC)},
            addressline={Avda. los Castros},
            city={Santander},
            postcode={39005},
            country={Spain}}

\affiliation[label5]{organization={Instituto de Microelectrónica de Barcelona (IMB-CNM-CSIC)},
            addressline={Cerdanyola del Vallès},
            city={Barcelona},
            postcode={08193},
            country={Spain}}



\begin{abstract}
Low Gain Avalanche Detectors built on n-type substrate (nLGADs) have been developed by IMB-CNM to enhance the detection of low-penetrating particles, with a wide range of applications from medicine, industry to synergies with developments for high-energy physics (HEP). In this work, irradiation effects on nLGADs were investigated through proton irradiation at the CERN PS-IRRAD facility up to proton fluences of \SI{1e14}{\per\centi\meter\squared}. Electrical characterization before and after irradiation reveals space charge sign inversion of the n-type bulk, leading to significant modifications in the depletion behavior and electric field distribution. Utilizing UV TCT and TPA-TCT measurements, the impact of irradiation on the electric fields and the gain are studied in more detail, confirming a change of sensor depletion and a reduced electric field in the gain layer. The results suggest that donor removal in nLGADs is stronger pronounced already at lower fluences compared to acceptor removal in traditional p-type LGADs. These findings provide not only first insights into the effects of irradiation on nLGADs but also contribute to the development of methods to quantify the donor removal.
\end{abstract}



\end{frontmatter}



\section{Introduction}
\label{intro}
Low Gain Avalanche Detectors (LGADs) are semiconductor detectors featuring a thin and highly doped gain layer for signal amplification and have proven their value in timing applications for high-energy physics (HEP) experiments. However, the performance of traditional p-type LGADs consisting of a p-type bulk and p-type gain layer is not optimized for the detection of low-penetrating particles such as soft X-rays, low-energy protons or UV photons. To overcome this limitation, n-type LGADs (nLGADs) were recently developed by IMB-CNM \cite{Villegas2023}. There are numerous applications for the detection of low-penetrating particles ranging from industry and medicine to scientific research. This paper presents first studies on the effects of irradiation damage on the novel nLGAD concept. One motivation to study the irradiation induced degradation of nLGADs is to probe their fundamental properties by using them as a tool for research on the physics of hole amplification and donor removal in the n-type gain layer. Current developments in HEP, such as the compensated LGAD concept \cite{Sola2024}, benefit from synergies with research on these topics, particularly regarding the fundamental understanding of donor removal in the n-type gain layer compared to the well-studied acceptor removal in traditional p-type LGADs. Furthermore, for the proposed usage in space or nuclear fusion experiments \cite{Han2022}, knowledge about the degradation of nLGADs in areas with high energy particle backgrounds can be of interest. 

\section{Materials and methods}
\subsection{Devices under test}
\begin{figure}[t]
\centering
\includegraphics[scale=0.4]{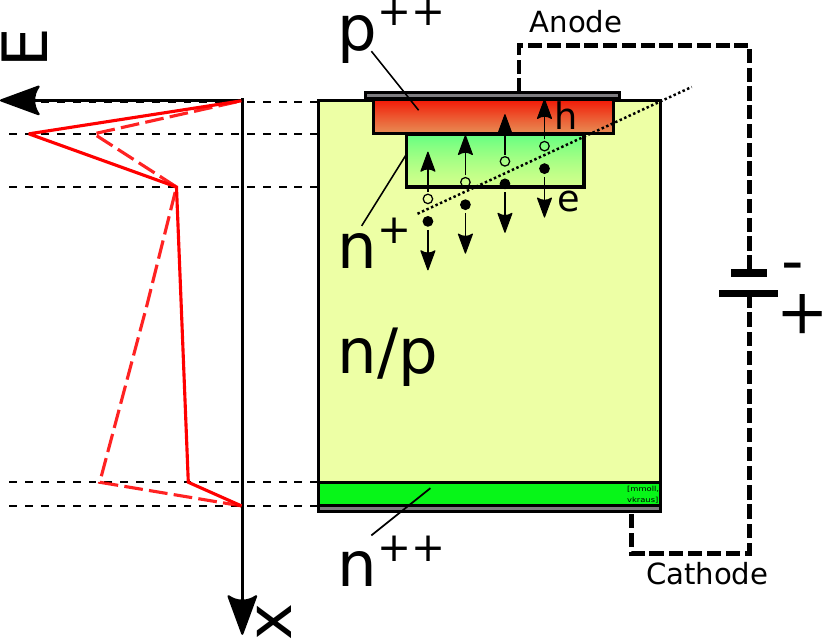}
\caption{Schematic view of an nLGAD implemented as $p^{++}-n^{+}-n$. The highly doped gain layer creates a high electric field (red, solid line) in order to exploit signal amplification for the detection of low-penetrating particles. With irradiation, the device undergoes space charge sign inversion which alters the electric field structure (red, dashed line).}
\label{fig1:nLGAD_sketch}
\end{figure}

The devices under test from IMB-CNM are implemented as $p^{++}-n^{+}-n$, i.e. with the same layers but inverted conductivity types compared to traditional p-type LGADs. A schematic view of the structure with a traversing particle can be seen in \cref{fig1:nLGAD_sketch}. The sketch also shows the advantage for low-penetrating particle detection in nLGADs, due to electrons initiating the impact ionization in the gain layer at a higher rate than holes. For the first nLGAD run R16375 from IMB-CNM, the devices under test are fabricated on n-type FZ wafers (CNM-4nLG1 technology). Spreading resistance profiling (SRP) measurements performed by CNM \cite{Villegas2025} show a low bulk doping < \SI{1e12}{atoms\per\centi\cubic\meter}, which means a high resistivity substrate. The thickness of the devices is \SI{275}{\micro\meter}, the square active area is \SI{1.3}{\milli\meter} $\times$ \SI{1.3}{\milli\meter} and features an opening in the topside metallization to enable laser characterization. 

\subsection{Irradiation}
The nLGADs together with corresponding reference PiNs from the same wafer were irradiated with \SI{23}{\giga\electronvolt} protons at the PS-IRRAD facility at CERN. The tested and here discussed proton fluences are \SI{1e12}{\per\centi\meter\squared}, \SI{1e13}{\per\centi\meter\squared}, \SI{2.4e13}{\per\centi\meter\squared}, \SI{6e13}{\per\centi\meter\squared}, \SI{1e14}{\per\centi\meter\squared}. After the irradiation, an annealing for \SI{4}{\minute} at \SI{80}{\degreeCelsius} was performed. The different measurement methods will be described in the corresponding sections. 

\section{Electrical characterization}
\begin{figure}[t]
\centering
\includegraphics[scale=0.48]{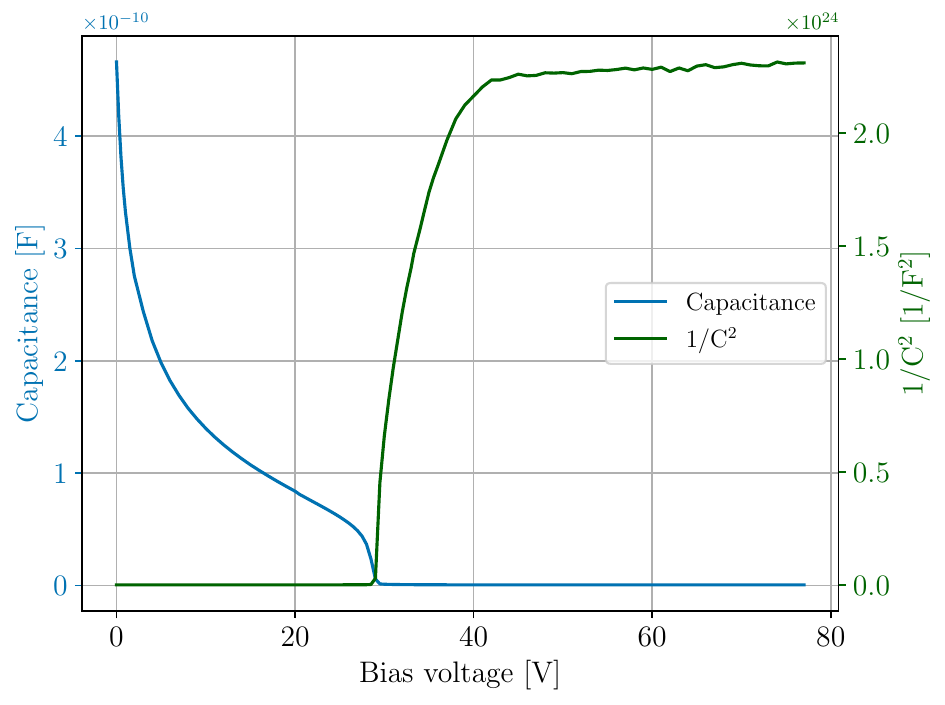}
\caption{Capacitance (blue, left axis) and inverse squared capacitance 1/C$^2$ (green, right axis) as a function of bias voltage for a CNM nLGAD before irradiation, measured at \SI{1}{kHz}. The steep decrease in capacitance and corresponding sharp rise in 1/C$^2$ reveal the depletion behavior of the sensor.}
\label{fig1:CV_before_irrad}
\end{figure}

The electrical characterization was performed at the SSD laboratory at CERN. A probe station to measure the current-voltage (IV) and capacitance-voltage (CV) characteristics before and after irradiation was used. The samples were placed directly on a temperature controlled chuck that was set to \SI{20}{\degreeCelsius}. During the electrical characterization, the guard ring was always grounded. The pad current as a function of the reverse bias voltage was measured before irradiation for all devices. The gain layer depletion voltage of the devices is measured to be \SI{\sim 28}{\volt} and the breakdown voltage lies within a range of \SI{200}{\volt} to \SI{250}{\volt}. The CV measurements before irradiation, performed at a frequency of \SI{1}{kHz} can be seen in \autoref{fig1:CV_before_irrad}. As evident from the additionally plotted 1/C$^2$ curve, full depletion of the sensor, as typical for LGADs, occurs shortly after the depletion of the gain layer at \SI{\sim 40}{V}.

\subsection{Indication for space charge sign inversion}
\begin{figure}[t]
    \centering
    \begin{subfigure}{0.48\textwidth}
        \centering
        \includegraphics[width=\textwidth]{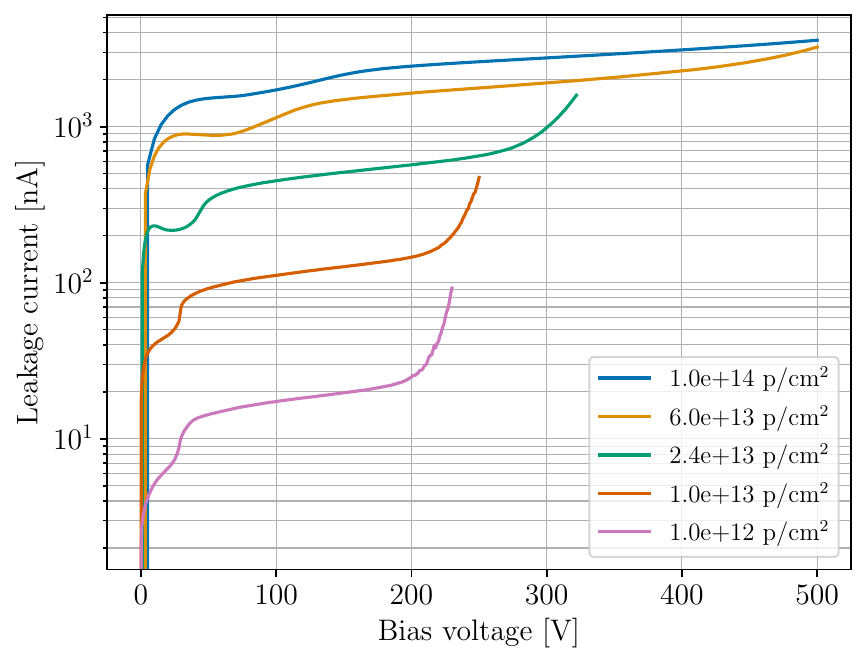}
        \caption{Leakage current versus bias voltage characteristics for increasing proton fluences. The typical shape of the curve changes and a temporary decrease in leakage current can be seen, which is associated with type inversion of the n-type silicon bulk.}
        \label{subfig:IV}
    \end{subfigure}
    \hfill
    \begin{subfigure}{0.48\textwidth}
        \centering
        \includegraphics[width=\textwidth]{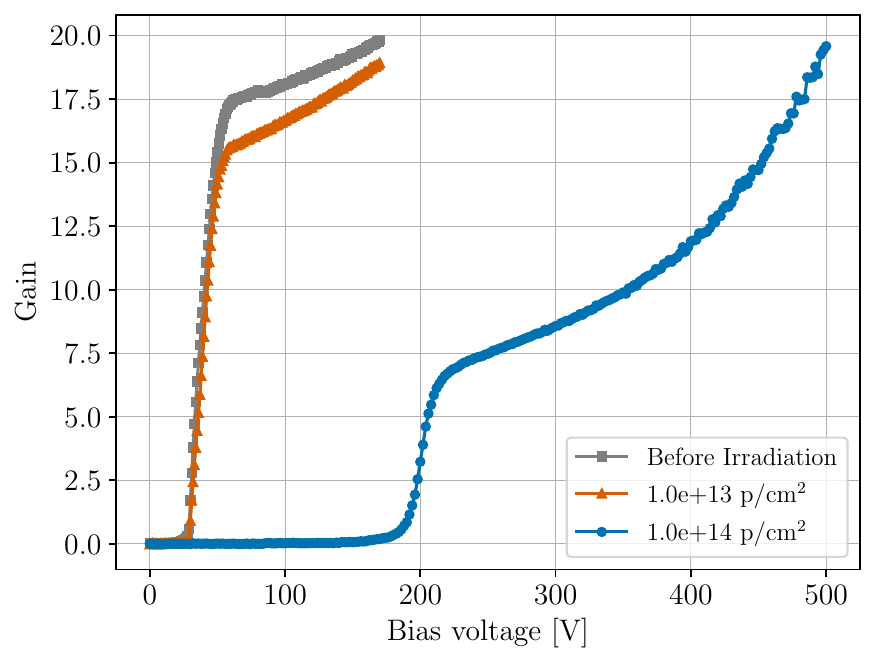}
        \caption{Gain measurements with an UV laser before and after proton irradiation.}
        \label{subfig:gain}
    \end{subfigure}
    \caption{Electrical and UV TCT measurements of proton irradiated nLGADs.}
    \label{fig2:IV_gain}
\end{figure}

After irradiation and the mentioned annealing of \SI{4}{\minute} at \SI{80}{\degreeCelsius}, the pad current as a function of the reverse bias voltage was measured again. The results for the five increasing proton fluences are displayed in \cref{subfig:IV}. In the shown plot, the results for the PiN diodes are not included, but they show a typical, increasing pad current with irradiation which is consistent with the proportionality described by the current related damage rate $\alpha$. 

For the two lowest fluences, the gain layer depletion, represented by the steep increase in the IV curve, and breakdown occur roughly at the same applied reverse bias voltage as before irradiation. With increasing proton fluences, the typical shape of the leakage current curve changes and a temporary decrease in leakage current can be seen roughly at the bias voltage were full depletion of the device occurs. This behavior can be associated with a space charge sign inversion of the n-type silicon bulk. Space charge sign inversion, also commonly referred to as type inversion in literature, is a known effect in n-type PiN diodes caused by a combination of donor removal and generation of acceptor-like states with irradiation \cite{Pitzl1992}. Numerous sources report space charge sign inversion in high resistivity n-type silicon diodes irradiated with various particles, most commonly neutrons and protons, at fluences between \SI{1e12}{\per\centi\meter\squared} and \SI{1e14}{\per\centi\meter\squared}. The most comparable irradiation conditions to the here presented study, involving proton exposure as described in \cite{Pitzl1992}, report space charge sign inversion at a fluence of \SI{1.5e13}{\per\centi\meter\squared}, which is in good agreement with the results presented for nLGADs. However, this effect and its impact on the electric field structures and thus on depletion and gain behavior have not been studied profoundly in LGAD structures before. Space charge sign inversion leads not only to a changed effective doping concentration, but also to modifications in the internal electric fields, which alter the standard depletion pattern. Rather than a uniform top-to-backside depletion, the process follows a more complex scheme after irradiation. When the guard ring is connected, charge carriers can drift towards it due to the altered field structures, temporarily reducing the measured leakage current before full depletion is established.
In addition, a shift of breakdown voltage to higher voltages for the largest fluences is visible, which is an indication for reduced impact ionization and therefore less gain after irradiation. The effect of degradation of the gain layer with irradiation is well-studied in traditional p-type LGADs and gets attributed to acceptor removal. The laser characterization described in the following will provide confirmation of space charge sign inversion and gain reduction in the irradiated nLGADs. 

\section{Laser characterization}
By using a pulsed laser light source, and measuring the current created by drifting access charge carriers, information about the changing gain and electric field structure in the nLGADs with irradiation can be obtained. 

\subsection{UV TCT}

For the laser measurements, the devices (nLGAD and reference PiN) are mounted on a passive PCB and read out using a Cividec C2 current amplifier. After the amplification stage, the signal gets digitized with an Agilent DSO 9254 Oscilloscope. To primarily trigger the avalanche mechanism through electron initiated impact ionization in the nLGADs, a UV laser with a wavelength of \SI{375}{\nano\meter} is used, as it has a low penetration depth in silicon, ensuring charge carrier generation near the surface. Initial tests with an IR laser have shown that the gain for this wavelength is very low, since the higher impact ionization rate of the electrons cannot be exploited if the light penetrates through the entire device. The used UV laser intensity is estimated to correspond to a deposited charge of \SI{\sim 59}{\femto\coulomb} in the devices. For all laser tests, the temperature for measurements after irradiation is set to \SI{-20}{\degreeCelsius}. 

\subsection{Evolution of gain with irradiation}

The gain can be expressed as the ratio between the collected charge (CC) of the nLGAD and reference PiN after full depletion as a function of the reverse bias voltage
\begin{equation}
Gain[V] = \frac{CC_{nLGAD}[V]}{CC_{PiN}[V \geq V_{FD}]} .
\end{equation}
The comparison of measured gain before irradiation and after exposure to proton fluences of \SI{1e13}{\per\centi\meter\squared} (before inversion of the n-type bulk to p-type is suspected) and \SI{1e14}{\per\centi\meter\squared} (after inversion) can be seen in \cref{subfig:gain}. The measurements show that the depletion behavior of the nLGAD changes after space charge sign inversion due to altered electric fields building up in the device. The device irradiated to \SI{1e14}{\per\centi\meter\squared}, represented by the blue curve in the plot, is no longer depleted from the top side via the gain layer, and only reaches full depletion after \SI{200}{\volt} compared to full depletion at \SI{\sim 40}{V} before irradiation. In addition, the gain is reduced after irradiation. Before irradiation, a gain value of 20 at a depletion voltage of \SI{160}{\volt} is reached. The same value of 20 is obtained only when a bias voltage of \SI{500}{\volt} is applied after a fluence of \SI{1e14}{\per\centi\meter\squared}. Extensive studies have established that irradiation induced performance degradation in the form of signal gain reduction is a well-known effect in traditional p-type LGADs, as for example reported in \cite{Kramberger2015} and \cite{Curras2023}. Comparisons to these sources suggest that the extent of gain loss in nLGADs seems to be higher for the relatively low fluences that were tested, which is an indication that donor removal is faster compared to acceptor removal.

\subsection{Two Photon Absorption - TCT}

The Two Photon Absorption – TCT setup at the CERN SSD laboratory was developed to obtain 3D spatial resolution in silicon sensors \cite{Wiehe2021}. A femtosecond laser with a wavelength in the quadratic absorption regime in silicon is focused to generate excess charge by two photon absorption in a small volume around the focal point of the laser. The setup features a FYLA LFC1500X fiber laser module as laser source with a wavelength of \SI{1550}{\nano\meter}. Downstream the laser source, the light is guided into the pulse management module, where the repetition rate can be adjusted, the pulse energy can be lowered to a given value using a neutral density filter (NDF), and a single photon absorption (SPA) energy reference is included. Afterwards, the beam traverses through open space inside a Faraday cage. The laser is guided through an objective onto the device under test for measurements. The PCB and readout configuration are identical to the above described UV TCT measurements. 

\subsection{Irradiation induced changes of the electric fields}

\begin{figure}[t]
    \centering
    \begin{subfigure}{0.48\textwidth}
        \centering
        \includegraphics[width=\textwidth]{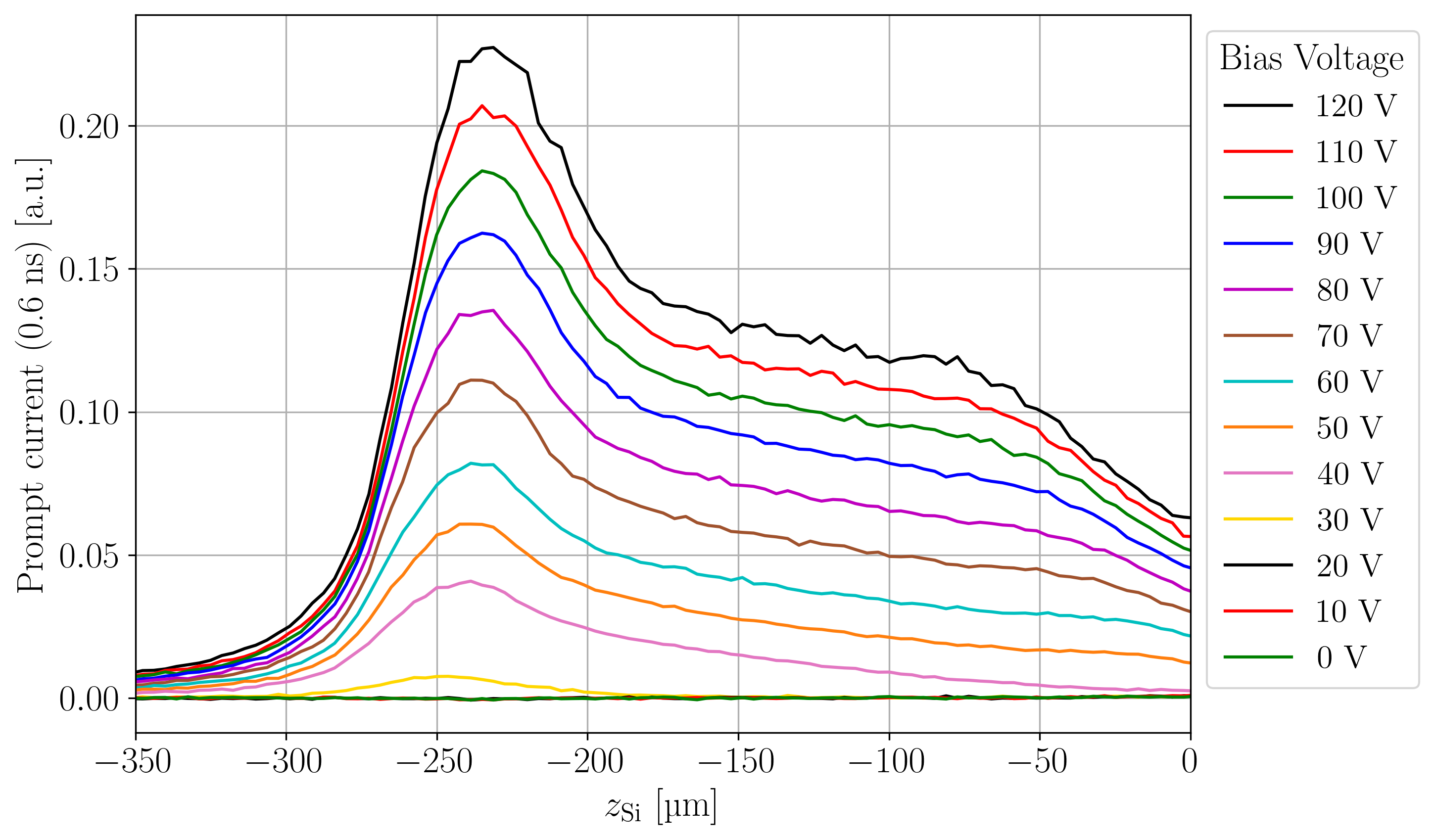}
        \caption{Electric field before irradiation.}
        \label{subfig:TPA_notIrrad}
    \end{subfigure}
    \hfill
    \begin{subfigure}{0.48\textwidth}
        \centering
        \includegraphics[width=\textwidth]{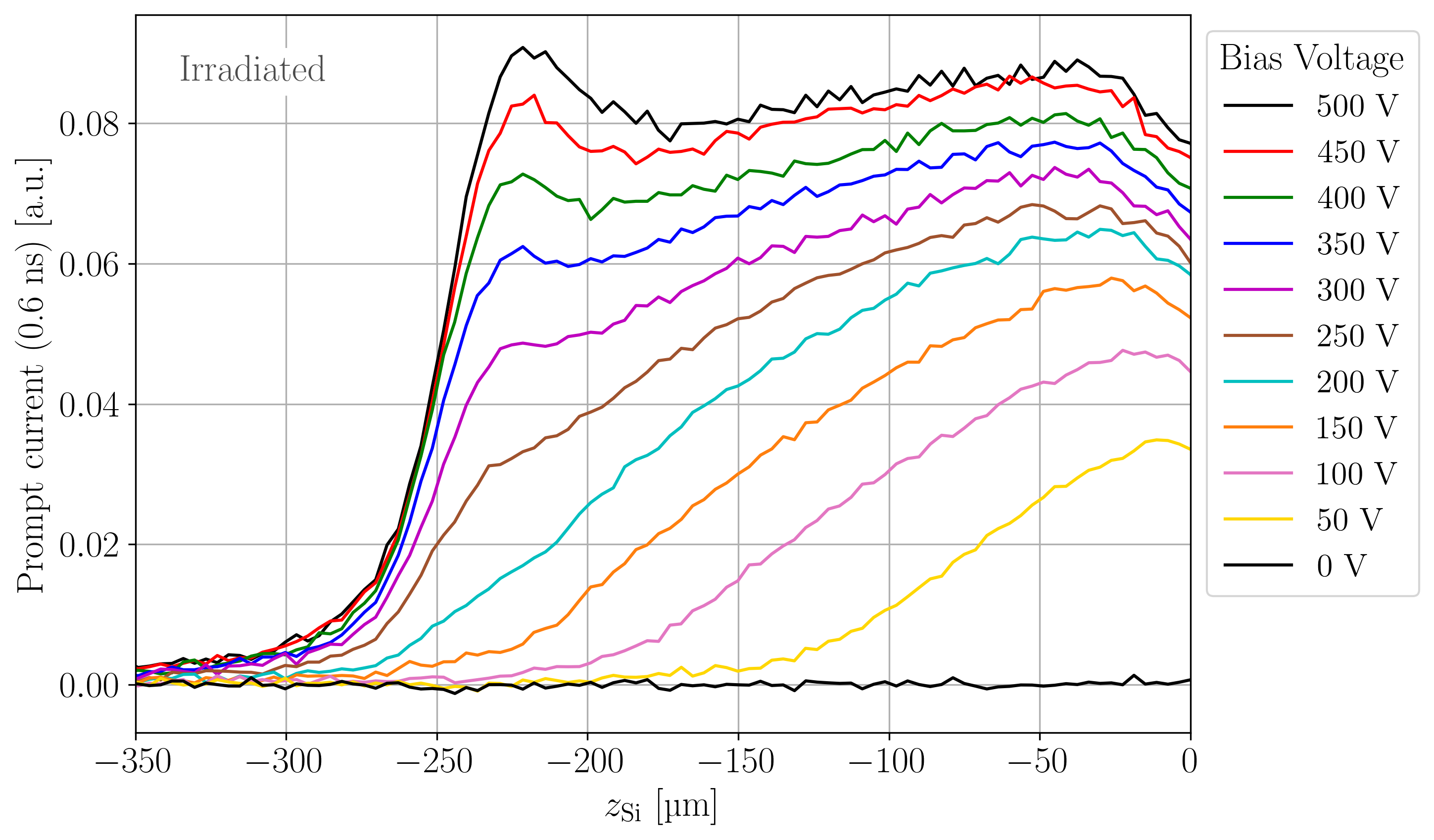}
        \caption{Electric field after exposure to a proton fluence of \SI{1e14}{\per\centi\meter\squared}.}
        \label{subfig:setup2}
    \end{subfigure}
    \caption{TPA-TCT measurements of the prompt current \SI{600}{\pico\second} after charge generation along the thickness of the nLGAD before and after irradiation. The prompt current represents the electric field in the device.}
    \label{fig3:TPA-TCT}
\end{figure}

With TPA-TCT, the prompt current directly after charge generation dependent on the position in the nLGAD can can be measured. Utilizing this prompt current method \cite{Kramberger2010}, the investigation of the electric field in the nLGAD before and after irradiation gets possible. The prompt current \SI{600}{\pico\second} after charge generation along the device thickness, representing the electric field, can be seen in \cref{fig3:TPA-TCT}. The two plots compare the electric fields before irradiation and after exposure to a proton fluence of \SI{1e14}{\per\centi\meter\squared}. Due to the changed depletion and breakdown behavior of the device after irradiation, the tested bias voltages differ. By comparing the electric fields before and after irradiation, the space charge sign inversion from an n-type bulk to a p-type bulk can be clearly confirmed. It can be seen that after the inversion the device starts to deplete from the backside ($z=0$ on the axis of the plot) and no longer from the front. The peak in the electric field where the gain layer is located, which is a characteristic of LGADs, is way less pronounced after irradiation.


\section{Conclusions}
\label{conclusion}
In this study, the irradiation tolerance of novel IMB-CNM nLGADs is investigated. The tested devices with a $p^{++}-n^{+}-n$ structure were irradiated with \SI{23}{\giga\electronvolt} protons at the CERN PS-IRRAD facility to fluences ranging from \SI{1e12}{\per\centi\meter\squared} up to \SI{1e14}{\per\centi\meter\squared}. Electrical characterization was performed at the SSD laboratory at CERN. Before irradiation a gain layer depletion voltage of \SI{\sim 28}{\volt}, full depletion at \SI{\sim 40}{V}, and breakdown voltages between \SI{\sim 200}{\volt} and \SI{\sim 250}{\volt} were measured.

Post-irradiation electrical characterization indicates space charge sign inversion of the n-type bulk, a phenomenon commonly observed in n-type PiN diodes. This inversion temporarily reduces the leakage current before full depletion when the guard ring is connected. The altered electric field distribution leads to a more complex depletion process, differing from the top-to-backside depletion before irradiation. Furthermore, with increasing fluence, a shift in breakdown voltage is observed, suggesting a reduction in maximum field strength and thus in impact ionization.

The change of gain and electric field structures with irradiation is further examined using UV TCT and TPA-TCT measurements. The results confirm the space charge sign inversion of the n-type bulk and therefore explain the shift in depletion behavior towards higher bias voltages. Before irradiation, depletion starts from the topside via the gain layer, whereas after space charge sign inversion, the nLGAD depletes from the backside. Additionally, a reduction in the characteristic gain layer electric field peak is observed. A comparison with traditional p-type LGADs suggests that donor removal in nLGADs occurs at lower fluences than acceptor removal in p-type LGADs, leading to a more rapid degradation of signal amplification.

The study of irradiation-induced degradation in nLGADs, particularly donor removal and its impact on the gain, remains challenging due to the complex electric field structures emerging after space charge sign inversion. This work provides the first experimental insights into these effects, laying the foundation for understanding this fundamental processes and developing a method to determine the donor removal coefficient. Future investigations will include the comparison of the presented study to neutron irradiation and low-energy proton irradiation. 

\section*{Acknowledgments}
This project has received funding from the European Union’s Horizon Europe Research and Innovation program under Grant Agreement No 101057511 (EURO-LABS). We are grateful to IMB-CNM for the samples, which have been funded by the Ministry of Science, Innovation and Universities (MICIU/AEI/10.13039/501100011033/). Grant references: PID2020-113705RB-C32, PID2023-148418NB-C42 and PDC2023-145925-C32.




\end{document}